\documentclass{article}
\usepackage[utf8]{inputenc}

\usepackage{authblk}
\usepackage{orcidlink}

\usepackage{hyperref}

\newcommand*{\email}[1]{\tt\href{mailto:#1}{#1}}

\title{Automatic Conversion of MiniZinc Programs to QUBO\footnote{The presented work was performed in the EniQmA project funded by the
Federal Ministry for Economic Affairs and Climate Action under the FKZ 01MQ22007A.}}

\author[1]{Armin Wolf\,\orcidlink{0000-0003-3940-0792}}
\author[2]{Cristian Grozea\,\orcidlink{0000-0001-6393-1919}}

\affil[1]{Fraunhofer FOKUS, Kaiserin-Augusta-Allee 31, D-10589 Berlin, Germany,
\email{armin.wolf@fokus.fraunhofer.de}}
\affil[2]{Fraunhofer FOKUS, Kaiserin-Augusta-Allee 31, D-10589 Berlin, Germany,
\email{cristian.grozea@fokus.fraunhofer.de}}

\date{Version of \today}

\usepackage{amsfonts}

\newcommand{\f}[1]{\mbox{\textsf{#1}}}

\newcommand{\fs}[1]{\mbox{\scriptsize\textsf{#1}}}

\begin{document}

\maketitle


\begin{abstract}
Obtaining  Quadratic Unconstrained Binary Optimisation models for various 
optimisation problems, in order to solve those on physical quantum computers (such 
as the the DWave annealers) is nowadays a lengthy and tedious process that requires 
one to remodel all problem variables as binary variables and squeeze the target 
function and the constraints into a single quadratic polynomial into these new 
variables. 

We report here on the basis of our automatic converter from MiniZinc to QUBO, which 
is able to process a large set of constraint optimisation and constraint satisfaction 
problems and turn them into equivalent QUBOs, effectively optimising the whole process. 
\end{abstract}

\section{Introduction}

Our work in the EniQmA project\footnote{see 
\href{https://www.eniqma-quantum.de/}{https://www.eniqma-quantum.de/}} aims to 
transform MiniZinc programs\footnote{see 
\href{https://www.minizinc.org/}{https://www.minizinc.org/}}
defining constraint satisfaction or optimisation problems (CSP/COP) into 
equivalent Quadratic Unconstrained Binary Optimisation (QUBO) problems.
The intention is to offer thereby a tool to support the solving of a large set
of satisfaction and optimisation problems by the use of Quantum Computing. 
In particular, to use Quantum Annealers beyond other solvers in the MiniZinc 
context.  

Thanks to the developers of the MiniZinc IDE we can focus on FlatZinc 
programs\footnote{The language FlatZinc is a proper subset of the language MiniZinc.} 
because any MiniZinc program can be transformed automatically by the MiniZinc compiler 
into an equivalent FlatZinc program.
A large set of FlatZinc programs are representing \emph{quadratic polynomial 
constraint optimisation problems} on finite domain integer variables. 
Therefore we focus in this document on some conceptional and theoretical work on 
transforming those problems into QUBO problems, although extension to floating point 
constants and variables is both possible and already working for us on limited 
experiments. 

The document is organised as follows: in the next section relevant related work 
is presented. Then, in Section~\ref{sec:3}, QUBO problems are introduced followed 
by the description of our aim, i.e. the transformation of CSP/COP into QUBO
problems, in Section~\ref{sec:4}. These problems are defined in Section~\ref{sec:5}, 
where it is also shown how they are transformed step-by-step into QUBO problems. 
The documents closes with conclusions and future work.

\section{Related Work}

The rich collection of Ising models for important NP-hard problems 
in~\cite{lucasIsingFormulationsMany2014a} gives an idea on how to model those
problems as QUBO problems, too, due to the fact that Ising models can be directly 
transformed into QUBO models, e.g. see \cite{karimiPracticalIntegertobinaryMapping2019}
for transforming spins into binary variables. In~\cite{lucasIsingFormulationsMany2014a}
an encoding of integer variables is presented that reduces the number of necessary 
binary variables from $O(N)$ to $O(\log N)$ where $N$ is the size of the domain of the
variable to be encoded. Both encodings are considered herein, because both have their
advantages in Quantum Annealing (cf.~\cite{karimiPracticalIntegertobinaryMapping2019}).

There exist also a bachelor thesis in the context of MiniZinc on transforming linear 
integer problems into QUBO~\cite{silvestreSolvingNPHardProblems2018}. However the 
transformation of integer linear programs into binary linear programs is only shown by 
example.

In~\cite{changIntegerProgrammingQuantum2020} a transformation of Integer Linear 
Programs (ILP) to QUBO problems is presented. However, it is ignored that the domains
of the variables must be finite and it assumes that the these domains are ranging
from $0$ to $2^k-1$ when encoding integer variables with binary variables. 

The Julia QUBO tools\footnote{available at \href{https://github.com/psrenergy/QUBOTools.jl}%
{https://github.com/psrenergy/QUBOTools.jl}} 
``\emph{\ldots implements codecs for QUBO (Quadratic Unconstrained Binary Optimisation)
instances. Its purpose is to provide fast and reliable conversion between common formats
used to represent such problems.}'' In particular, the aim is to transform MiniZinc
programs into QUBO problems. Currently, only a small subset of the MiniZinc language 
is supported by this tool.

ToQUBO.jl~\cite{toqubo:2023} is a Julia package for transforming 
optimisation problems formulated in JuMP (Julia embedded mathematical 
optimisation language) into QUBO instances. This Julia package is built on top
of the Julia QUBO tools.

\section{QUBO Problems} \label{sec:3}

A \emph{quadratic unconstrained binary optimisation (QUBO) problem} is a combinatorial 
optimisation problem: Let some binary variables $x = x_1, \ldots, x_n$ and a 
rational-valued matrix~$Q \in \mathbb{Q}^{n\times n}$ be given where each value 
$Q_{ij}$ defines a weight for the pair of indices $i,j = 1,\dots ,n$. Then the 
according \emph{QUBO problem} is to find values~$x_{\min} \in \{0,1\}^n$ such that
\begin{eqnarray}
    x_{\min} = \arg\min_{x \in \{0,1\}^n} x^{T} Q x \enspace,
\end{eqnarray}
i.e. the quadratic term~$x^{T} Q x$ 
has to be minimised. 
We assume that the QUBO  problem is \emph{normalised}, i.e. the matrix~$Q$ is an upper 
triangular matrix such that $Q_{ij} = 0$ for $j < i$ and
\begin{eqnarray}
   x^{T} Q x & = &  \sum_{i=1}^n \sum_{j=i}^n Q_{ij} \cdot x_i \cdot x_j
\end{eqnarray}
Please note that there is an alternative but equivalent representation of a QUBO problem
where each binary product~$x_i^2$ is equivalently replaced by $x_i$:
\begin{eqnarray}
    x^{T} Q x & = & \sum_{i=1}^n Q_{ii} \cdot x_i + \sum_{i=1}^n \sum_{j=i+1}^n 
    Q_{ij} \cdot x_i \cdot x_j \enspace .
\end{eqnarray}

\section{Our Aim} \label{sec:4}

Our aim is the transformation of any finite domain, integral, linear, quadratic or 
even polynomial constraint (optimisation) problem into an \emph{equivalent} QUBO 
problem with a rather small number of binary variables due to the fact that current 
Quantum Annealers and Quantum Computers only support a rather small number of qubits
in the so-called NISQ era.
Furthermore, our aim is to maintain the substitutions that are necessary to transfer
the solutions of the QUBO problems back into the according solutions of the original 
problem.
  
\section{Finite Domain Quadratic Integer Programs} \label{sec:5}

In the following we restrict ourselves to \emph{Finite Domain Quadratic Integer Programs 
(QIP(FD))}, e.g. semantically equivalent to plain integer {\tt FlatZinc} programs. 
In detail, we focus on integer optimisation problems of the form
\begin{eqnarray}
    x_{\min} = \arg \min_{x} \left(\sum_{i=1}^n g_i \cdot x_i\right) \label{eqn:orgObj}
\end{eqnarray}
subject to 
\begin{eqnarray}
    \sum_{i=1}^n a_{j,i} \cdot x_i + \sum_{k=1}^m b_{j,k} \cdot y_k + c_j \le 0 
    & & \mbox{for $j = 1, \ldots, p$} \label{eqn:orgIneqs} \\
    \sum_{i=1}^n d_{j,i} \cdot x_i + \sum_{k=1}^m e_{j,k} \cdot y_k + f_j = 0 
    & & \mbox{for $j = p+1, \ldots, q$} \\
    y_k = z_v \cdot z_w  
    & & \mbox{$z_v, z_w \in \{x_1, \ldots, x_n, y_1, \ldots, y_{k-1}\}$} \nonumber \\
    & & \mbox{for $k = 1 \ldots, m$} \label{eqn:orgProd}\\
    x_i \in D(x_i) \subset \mathbb{Z} & & \mbox{for $i = 1, \ldots, n$} \\
    y_k \in D(y_k) \subset \mathbb{Z} & & \mbox{for $k = 1, \ldots, m$} \enspace,
    \label{eqn:orgLast}
\end{eqnarray}
where $g_i, a_{j,i}, b_{j,k}, c_j,  d_{j,i}, e_{j,k}, f_j$ are 
rational values, $x_i, y_k$ are uniquely defined variables with 
non-empty finite domains $D(x_i)$ resp. $D(y_k)$. It should be noted that any 
polynomial integer program over finite domains can be transformed into a 
QIP(FD).\footnote{We do not consider such transformations. They will be performed 
by the MiniZinc tools.}

In the special case that the variable products -- Equations~(\ref{eqn:orgProd}) -- are
missing we have Finite Domain Linear Integer Programs, LIP(FD). 

We assume that the finite domains of the variables are \emph{bounds-consistent integer 
interval}, e.g. according to the pruning rules defined 
in~\cite{schulteWhenBoundsDomain2001}, Figure~1.

We further assume that there might be some \emph{substitutions}~$S$ consisting not
only for the variables~$x_i$ or~$y_k$, e.g. $s := t(s_1, \ldots, s_l)$ where $s$ is 
the substituted variable and $t(s_1, \ldots, s_l)$ is a linear or affine term over some 
other variables~$s_1, \ldots, s_l$ --- maybe including other substituted variables, too 
--- such that the value of~$s$ can be computed by using the values of the variables 
$s_1, \ldots, s_l$. Therefore the substitutions must construct a \emph{forest of 
finite trees} in a graph-theoretical sense where the variables are the \emph{nodes} 
of the trees and binary variables of the resulting QUBO problem are \emph{leaves} 
of the trees. Initially the set of substitutions~$S$ is in general empty.
There might be one exception: if there were any variables~$x$ with domain $D(x)=\{v\}$ 
which were eliminated in advance. Then~$S$ contains substitutions~$x = v$. Thus it is
assumed that for each variable~$x$ it holds that it is not determined in advance, i.e. 
$|D(x)| > 1$. 

Due to the fact that at the end of the transformation of a QIP(FD) into a QUBO problem 
there will be one objective to be minimised, we assume that the output variables 
$x_1, \ldots, x_n$ are stored in the set~$O = \{x_1, \ldots, x_n\}$ and that the 
Equation~(\ref{eqn:orgObj}) is simplified to $\min(\sum_{i=1}^n g_i \cdot x_i)$.

\subsection{Eliminating Inequalities by the Use of Slack Variables}\label{sec:slack}

For the transformation of a QIP(FD) into a QUBO problem we have to get rid of
the inequalities by replacing them by equivalent equations using \emph{slack 
variables}. For example the inequality $3\cdot x - 2\cdot y \le 0$ with $x \in 
[0, 1],$ and $y \in [0, 2]$ can be equivalently replaced by $3 \cdot x - 2 \cdot 
y = z$ where $z \le 0$ or more precisely by $3 \cdot x - 2 \cdot y - z = 0$ where 
$z \in [-4, 0]$. However, we can set $z' = -z$ with $z' \in [0, 4]$ such that the 
domain of the additional variable becomes `canonical', i.e. starting at zero.

In order to replace the Inequalities~(\ref{eqn:orgIneqs})  by equations adequately let
\begin{eqnarray}
    l_j & = & c_j + \sum_{i=1 \land a_{ji} < 0}^n a_{ji} \cdot\max(D(x_i)) 
        + \sum_{i=1 \land a_{ji} > 0}^n a_{ji} \cdot\min(D(x_i)) \nonumber \\
        &   & + \sum_{k=1 \land b_{jk} < 0}^m b_{jk} \cdot\max(D(y_k)) 
        + \sum_{k=1 \land b_{jk} > 0}^m b_{jk} \cdot\min(D(y_k))  \\
    u_j & = & c_j + \sum_{i=1 \land a_{ji} < 0}^n a_{ji} \cdot\min(D(x_i)) 
        + \sum_{i=1 \land a_{ji} > 0}^n a_{ji} \cdot\max(D(x_i)) \nonumber \\
        &   & + \sum_{k=1 \land b_{jk} < 0}^m b_{jk} \cdot\min(D(y_k)) 
        + \sum_{k=1 \land b_{jk} > 0}^m b_{jk} \cdot\max(D(y_k))
\end{eqnarray}
for $j= 1, \ldots, p$ be the lower and upper bounds of the left-hand sides of the 
Inequalities~(\ref{eqn:orgIneqs}). 

If $l_j > 0$ then the according inequality cannot be satisfied, i.e. the QIP(FD) is 
inconsistent and has not any solution. If $u_j \le 0$ then the according inequality is 
always satisfied and can be omitted --- it holds that $0 \le l_j \le u_j \le 0$ and 
finally $l_j = u_j = 0$.
In all other cases we can reformulate the QIP(FD) equivalently as
\begin{eqnarray}
    \min \left(\sum_{i=1}^n g_i \cdot x_i\right)
\end{eqnarray}
subject to 
\begin{eqnarray}
    \sum_{i=1}^n a_{j,i} \cdot x_i + \sum_{k=1}^m b_{j,k} \cdot y_k + s_j + c_j = 0 
    & & \mbox{for $j = 1, \ldots, p$} \nonumber \\[-3ex] 
    & & \mbox{and if $l_j \le 0 < u_j$} \\
    \sum_{i=1}^n a_{j,i} \cdot x_i + \sum_{k=1}^m b_{j,k} \cdot y_k + c_j = 0 
    & & \mbox{for $j = 1, \ldots, p$} \nonumber \\[-3ex] 
    & & \mbox{and if $l_j = 0 = u_j$}
\end{eqnarray}
\begin{eqnarray}
    \sum_{i=1}^n d_{j,i} \cdot x_i + \sum_{k=1}^m e_{j,k} \cdot y_k + f_j = 0 
    & & \mbox{for $j = p+1, \ldots, q$} \\
    y_k = z_v \cdot z_w  
    & & \mbox{$z_v, z_w \in \{x_1, \ldots, x_n, y_1, \ldots, y_{k-1}\}$} \nonumber \\
    & & \mbox{for $k = 1 \ldots, m$} \\
    x_i \in D(x_i) \subset \mathbb{Z} & & \mbox{for $i = 1, \ldots, n$} \\
    y_k \in D(y_k) \subset \mathbb{Z} & & \mbox{for $k = 1, \ldots, m$} \\
    s_j \in [0, -\lceil l_j \rceil] \subset \mathbb{Z} & & \mbox{for $j = 1, \ldots, p$}
    \nonumber \\
    & & \mbox{and if $l_j < 0 < u_j$}
\end{eqnarray}
This transformation introduces at most~$p$ new variables with canonical lower bounds (cf.
Section~\ref{sec:canonical}) but not any substitution. 

\subsection{Canonical Lower Bounds of Variable Domains}\label{sec:canonical}

The formulation of a QIP(FD) as QUBO problem even the encoding of the integer variables
with binary variables becomes simpler if the domains  of the integer variables start 
canonically with zero. 

Let a QIP(FD) defined by the Equations~(\ref{eqn:orgObj}) -- (\ref{eqn:orgLast}) be 
given, where $p=0$ holds, i.e. there are not any inequalities: the 
Inequalities~(\ref{eqn:orgIneqs}) are missing. Then we can reformulate the 
QIP(FD) equivalently with canonical domains. Therefore, we substitute any integer 
variable~$x_l$ with $l \in \{1, \ldots, n\}, \min(D(x_l) \ne 0$ by $x'_l + \min(D(x_l))$ 
such that the resulting equivalent QIP(FD) is
\begin{eqnarray*}
    \min \left(\sum_{i=1}^n g_i \cdot z_i\right)
    & & \mbox{where $z_l \equiv x'_l$ if $x_l$ is substituted}  \nonumber \\[-3ex]
    & & \mbox{and $z_l \equiv x_l$ otherwise.}
\end{eqnarray*}
subject to 
\begin{eqnarray*}
    \sum_{i=1}^n d_{j,i} \cdot z_i
    + \sum_{k=1}^m e_{j,k} \cdot y_k + f_j + a_{j,l} \cdot \min(D(x_l)) = 0
    & \mbox{for $j = 1, \ldots, q$} & \nonumber \\[-6ex]
\end{eqnarray*}
\begin{eqnarray*}
    y_k  = z_v \cdot z_w 
    & & \mbox{$z_v, z_w \in \{x_1, \ldots, x_n, y_1, \ldots, y_{k-1}\}\setminus\{x_l\}$} 
    \nonumber \\
    & & \mbox{for $k = 1 \ldots, m$} \\ 
    y_k - y_{m+1} - \min(D(x_l)) \cdot z_w  = 0 
    & &  \mbox{if $z_v = x_l$} \nonumber \\
    y_{m+1}  = x'_l \cdot z_w  
    & & \mbox{and $z_w \in \{x_1, \ldots, x_n, y_1, \ldots, y_{k-1}\}\setminus\{x_l\}$} 
    \nonumber \\
    & & \mbox{for $k = 1 \ldots, m$}
\end{eqnarray*}
\begin{eqnarray*}
    y_k - y_{m+1} - 2\cdot\min(D(x_l))\cdot y_{m+1}  - \min(D(x_l))^2 = 0 
    & &  \mbox{if $z_v = z_w = x_l$} \nonumber \\
    y_{m+1} = {x'_l}^2 & & \mbox{for $k = 1 \ldots, m$}
\end{eqnarray*}
\begin{eqnarray*}
    z_i \in D(z_i) \subset \mathbb{Z} & & \mbox{for $i = 1, \ldots, n$} \\
    y_k \in D(y_k) \subset \mathbb{Z} & & \mbox{for $k = 1, \ldots, m$} 
\end{eqnarray*}
where a new variable $y_{m+1}$ is added if necessary and its domain~$D(y_{m+1})$ is
computed following the rules of interval arithmetic.

There are at most two new variables introduced and the set of substitutions~$S$ is
extended accordingly:
\begin{eqnarray}
    S & := & S \cup \{x_l := x'_l + \min(D(x_l))\} \enspace.
\end{eqnarray}
The resulting integer optimisation problem is a QIP(FD) without inequalities but
with one more variable having a canonical domain. There are at most two new variables
added and at most one additional linear equation. The number of products is not changing.

The direct substitution of any~$y_l$ with $l \in \{1, \ldots, m\}$ can be avoided if the
variables $z_v, z_w \in \{x_1, \ldots,x_n\}$ with $y_1 = z_v \cdot z_w$ are substituted.
This results in a QIP(FD) with the same number of products, where the product 
$y_1 = z_v \cdot z_w$ is replaced by a product of new variables and the variable~$y_1$ 
will be part of a linear equation and no longer involved in a product, i.e. it becomes
an `$x_l$' in the next step which can be substituted if necessary.
The most recently introduced variable~$y_{m+2}$ is then the product of two variables 
having domains where their minima are zero. Thus, the minimum of the domain~$D(y_{m+2})$ 
is zero, too. This procedure can be repeated for $y_2, \ldots, y_m$ until all variables 
in products have domains with zero as minimum. 

It is strongly recommended to apply this kind of normalisation for any variable~$x_l$
with $\min(D(x_l)) \ne 0$ and $\max(D(x_l)) = min(D(x_l))+1$ because then the domain 
of~$x'_l$ becomes $\{0,1\}$ and the resulting QIP(FD) will contain a binary 
variable~$x'_l$ instead of an integer variable~$x_l$.

Altogether, this kind of normalisation requires only a finite number of transformation
steps until all domains are canonical.

\subsection{Computing a Weighting Factor for Penalty Conditions}\label{sec:factors}

The formulation of a QIP(FD) as an equivalent QUBO problem requires the transformation 
of the constraining linear equations into (linear or quadratic) penalty 
terms~$t_1, \ldots, t_p$ over binary variables of an extended objective function 
such that these terms will be zero if the according equations are satisfied and 
greater than zero if not. Thus we are looking for penalty factors~$C_1, \ldots, C_p$ 
of these penalty terms such that it will hold
\begin{eqnarray} 
    \sum_{i=1}^n g_i \cdot x_i + \sum_{j=1}^p C_j \cdot t_j > 
    \max_{y_1, \ldots, y_n} \sum_{i=1}^n g_i \cdot y_i & & \mbox{if there is a 
    $t_j > 0$,}
\end{eqnarray}
i.e. any violation of a constraint will result in objective value that is worse than
the worst value of the original objective. For each penalty term~$t_j$ there is an 
$\varepsilon_j > 0$ such that $t_j \ge \varepsilon_j$ if $t_j > 0$ holds --- due to
the finite domains of the variables --- i.e. if the according equation is violated. 
Then it holds
\begin{eqnarray} 
    \lefteqn{\sum_{j=1}^p C_j \cdot t_j > C_j \cdot t_j \ge C_j \cdot \varepsilon_j} \\
    & > &  \max_{y_1, \ldots, y_n} \sum_{i=1}^n g_i \cdot y_i 
    - \min_{y_1, \ldots, y_n} \sum_{i=1}^n g_i \cdot y_i \\
    & \ge & \max_{y_1, \ldots, y_n} \sum_{i=1}^n g_i \cdot y_i 
    - \sum_{i=1}^n g_i \cdot x_i  \\
    & & \mbox{if $t_j > 0$.}
\end{eqnarray}
This is the case if
\begin{eqnarray}
    C_j & > & \frac{\max_{y_1, \ldots, y_n} \sum_{i=1}^n g_i \cdot y_i - \min_{y_1, 
    \ldots, y_n} \sum_{i=1}^n g_i \cdot y_i}{\varepsilon_j}
\end{eqnarray}
holds where 
\begin{eqnarray*}
    \max_{y_1, \ldots, y_n} \sum_{i=1}^n g_i \cdot y_i 
    & = & \sum_{i=1 \land g_i < 0}^n g_i \cdot\min(D(x_i)) + 
    \sum_{i=1 \land g_i > 0}^n g_i \cdot\max(D(x_i)) \\
    \min_{y_1, \ldots, y_n} \sum_{i=1}^n g_i \cdot y_i 
    & = & \sum_{i=1 \land g_i < 0}^n g_i \cdot\max(D(x_i)) + 
    \sum_{i=1 \land g_i > 0}^n g_i \cdot\min(D(x_i)) \enspace.
\end{eqnarray*}

In general it can be inefficient to compute these $\varepsilon_j > 0$ for 
each linear or quadratic penalty term~$t_j$. Due to the fact that $t = \sum_{i=1}^n c_i 
\cdot x_i + d$ or $t = (\sum_{i=1}^n c_i \cdot x_i + d)^2$ with rational 
coefficients~$c_i$ and rational offset~$d$ holds then for the greatest denominator~$g 
\in \mathbb{N}$ of~$c_i$ and~$d$ it holds that $|\sum_{i=1}^n g\cdot c_i \cdot x_i + g 
\cdot d| \ge 1$ if $\sum_{i=1}^n g\cdot c_i \cdot x_i + g \cdot d  \ne 0$. This means
that if we consider $t = \sum_{i=1}^n c_i \cdot x_i + d$ or $t = (\sum_{i=1}^n c_i \cdot 
x_i + d)^2$ with $c_i \in \mathbb{Z}$ and $d \in \mathbb{Z}$ then let $\varepsilon = 
\varepsilon_j = 1$. We observe that in this case the $C_j$ are all the same for the 
terms~$t_j$, i.e. $C_j = C$. This determination of a common weighting factor~$C$ is only
one alternative and might be not the best, even if the values are rather big. However,
there are only heuristics for the choice of those weights which leaves space for further 
research.

\subsection{One Hot Encoding of QIP(FD)}\label{sec:oneHotEncoding}

Let a QIP(FD) defined by the Equations~(\ref{eqn:orgObj}) -- (\ref{eqn:orgLast}) be 
given, where $p=0$ holds, i.e. there are not any inequalities: the 
Inequalities~(\ref{eqn:orgIneqs}) are missing. We assume that there is integer 
variable~$x_l$ which is not the result of a variable product but with 
$D(x_l) = \{d_1, \ldots, d_h\}, h > 2$ then we can substitute $x_l$ by
$\sum_{p=1}^h d_p \cdot x_l^{(p)}$ when the equation 
\begin{eqnarray}
    \sum_{p=1}^h x_l^{(h)} & = & 1 
\end{eqnarray} 
is added and for the binary variables~$x_l^{(p)}, p = 1, \ldots, h$ it holds that
$D(x_l^{(p)}) = \{0,1\}$. 
Then the resulting QIP(FD) is
\begin{eqnarray}
    \min \left(\sum_{i=1 \land i \ne l}^n g_i \cdot x_i + 
    \sum_{p=1}^h g_l \cdot d_p \cdot x_l^{(p)}\right)
\end{eqnarray}
subject to 
\begin{eqnarray}
    \lefteqn{\sum_{i=1 \land i \ne l}^n d_{j,i} \cdot x_i 
    + \sum_{p=1}^h d_{j,l} \cdot d_p \cdot x_l^{(p)}} \\
    & & ~~~~~~~~~~~~~+~\sum_{k=1}^m e_{j,k} \cdot y_k + f_j = 0 
    \qquad \mbox{for $j = 1, \ldots, q$} \\
    \lefteqn{\sum_{p=1} x_l^{(p)} -1 = 0}
\end{eqnarray}
\begin{eqnarray}
    y_k  = z_v \cdot z_w 
    & & \mbox{$z_v, z_w \in \{x_1, \ldots, x_n, y_1, \ldots, y_{k-1}\}\setminus\{x_l\}$} 
    \nonumber \\
    & & \mbox{for $k = 1 \ldots, m$} \\ 
    y_k - \sum_{p=1}^h d_p \cdot y_{m+p} = 0 
    & &  \mbox{if $z_v = x_l$ and $z_w \in \{x_1, \ldots, x_n, y_1, \ldots, 
    y_{k-1}\}\setminus\{x_l\}$} \nonumber \\[-2ex]
    y_{m+p}  = x_l^{(d_p)} \cdot z_w 
    & & \mbox{for $p = 1 \ldots, h$ and for $k = 1 \ldots, m$}
\end{eqnarray}
\begin{eqnarray}
    y_k - \sum_{p=1}^h d_p^2 \cdot y_{m+p} - \sum_{p=1}^h \sum_{r=p+1}^h 
    2 \cdot d_p \cdot d_r \cdot y_{m+h+s} = 0 & \mbox{if $z_v = z_w = x_l$} & \nonumber 
    \\[-6ex]
\end{eqnarray}
\begin{eqnarray}
    y_{m+p} & = & (x_l^{(p)})^2 \quad \mbox{for $p=1, \ldots, h$} \\ 
    y_{m+h+s} & = & x_l^{(p)} \cdot x_l^{(r)} \quad
    \mbox{where $s = \sum_{\phi=1}^p \sum_{\rho=\phi+1}^r 1$}
\end{eqnarray}
\begin{eqnarray}
    x_i \in D(x_i) \subset \mathbb{Z} & & \mbox{for $i = 1, \ldots, l-1, l+1, \ldots n$} 
    \\
    y_k \in D(y_k) \subset \mathbb{Z} & & \mbox{for $k = 1, \ldots, m (+ h 
    (, \ldots,m+h+h(h-1)/2))$} \\
    x_l^{(p)} \in \{0,1\} & & \mbox{for $p = 1, \ldots, h$}
\end{eqnarray}
The set of substitutions~$S$ is
extended accordingly:
\begin{eqnarray}
    S & := & S \cup \{x_l := \sum_{p=1}^h d_p \cdot x_l^{(p)} \} \enspace.
\end{eqnarray}

The direct substitution of any~$y_l$ with $l \in \{1, \ldots, m\}$ can be avoided if the
variables $z_v, z_w \in \{x_1, \ldots,x_n\}$ with $y_1 = z_v \cdot z_w$ are substituted; 
see Section~\ref{sec:canonical} for details.

Finally, there are $h$ new binary variables introduced and either zero or $h$ or at most 
$h+h(h-1)/2$ new integer variables. It is recommended to substitute integer variables 
with rather small domains, even if the substituted variable occurs as a multiplier in 
a product. Then the number of integer variables increase significantly and those 
variables have be substituted in subsequent steps, too  --- either with one hot 
encoding or binary encoding.

Altogether, this encoding requires only a finite number of transformation
steps until all variables are binary.

\subsection{Binary Encoding of QIP(FD)}\label{sec:binEncoding}

Let a QIP(FD) defined by the Equations~(\ref{eqn:orgObj}) -- (\ref{eqn:orgLast}) be 
given, where $p=0$ holds, i.e. there are not any inequalities: the 
Inequalities~(\ref{eqn:orgIneqs}) are missing. We assume that there is integer 
variable~$x_l$ with $D(x_l) = [0,\max(D(x_l))], \max(D(x_l)) > 1$ then we can substitute 
$x_l$ by of sum of powers of two. A straightforward usual binary encoding could 
have been used, however, this would have required adding inequalities for 
enforcing the bounds of the variables with associated slack variables. Therefore
we prefer using self-bounding binary encoding which cover the domains automatically
without requiring additional slack variables. We use those two alternative encoding 
methods on $x \equiv x_l$:
\begin{enumerate}
    \item If $\max(D(x)) = 2^{r_x+1}-1$ holds we define new additional binary variables 
    $\f{var}(x) = \{x_0, \ldots x_{r_x}\}$ and replace $x$ by the sum
        \begin{eqnarray}
            \f{bin}(x) & = &  \sum_{s=0}^{r_x} 2^s \cdot x_s\enspace.
        \end{eqnarray}
    \item If $2^{r_x} \le \max(D_x) < 2^{r_x+1}-1$ holds we define
        new additional binary variables $\f{var}(x) = \{x_0, \ldots x_{r_x-1}\} \cup 
        \f{var}(x')$
        where~$x'$ is an `intermediate' integer variable with `intermediate' domain 
        \begin{eqnarray}
            D(x') = [0,\max(D(x)) - 2^{r_x}-1] \enspace,
        \end{eqnarray}
        and replace $x$ by the sum
        \begin{eqnarray}
            \f{bin}(x) & = & \sum_{s=0}^{r_x-1} 2^s \cdot x_s + \f{bin}(x') \enspace,
        \end{eqnarray}
        and further apply these rules recursively on the `intermediate' integer 
        variable~$x'$ while replacing this `intermediate' variable immediately within 
        this sum.
\end{enumerate}
Alternatively (cf.~~\cite{lucasIsingFormulationsMany2014a}), we can replace the second 
rule by
\begin{enumerate}
    \item[2'.] If $2^{r_x} \le \max(D_x) < 2^{r_x+1}-1$ holds we define
        new additional binary variables $\f{var}(x) = \{x_0, \ldots x_{r_x-1}, x_{r_x}\}$
        where~$x'$ is an `intermediate' integer variable with `intermediate' domain 
        \begin{eqnarray}
            D(x') = [0,\max(D(x)) - 2^{r_x}-1] \enspace,
        \end{eqnarray}
        and replace $x$ by the sum
        \begin{eqnarray}
            \f{bin}(x) & = & \sum_{s=0}^{r_x-1} 2^s \cdot x_k 
            + (\max(D(x)) - 2^{r_x}-1) \cdot x_{r_x} \enspace.
        \end{eqnarray}
\end{enumerate}

In both cases the application of the rules to an integer variable~$x_l$ with a non-binary
domain and which is not the result of variable product results in
\begin{eqnarray}
    \f{bin}(x_l) & = &  \sum_{s=0}^{r_{x_l}'} v_s \cdot x_{l_s} \\
    \f{var}(x_l) & = & \{x_{l_0}, \ldots, x_{l_{r_{x_l}'}} \}
\end{eqnarray}
and the resulting QIP(FD) is
\begin{eqnarray}
    \min \left(\sum_{i=1 \land i \ne l}^n g_i \cdot x_i + 
    \sum_{s=0}^{r_{x_l}'} v_s \cdot x_{l_s}\right)
\end{eqnarray}
subject to 
\begin{eqnarray}
    \lefteqn{\sum_{i=1 \land i \ne l}^n d_{j,i} \cdot x_i 
    + \sum_{s=0}^{r_{x_l}'} v_s \cdot x_{l_s}} \\
    & & ~~~~~~~~~~~~~+~\sum_{k=1}^m e_{j,k} \cdot y_k + f_j = 0 
    \qquad \mbox{for $j = 1, \ldots, q$}
\end{eqnarray}
\begin{eqnarray}
    y_k  = z_v \cdot z_w 
    & & \mbox{$z_v, z_w \in \{x_1, \ldots, x_n, y_1, \ldots, y_{k-1}\}\setminus\{x_l\}$} 
    \nonumber \\
    & & \mbox{for $k = 1 \ldots, m$} \\ 
    y_k - \sum_{s=0}^{r_{x_l}'} v_s \cdot \cdot y_{m+s} = 0 
    & & \mbox{if $z_v = x_l$} \\[-2ex]
    & & \mbox{and $z_w \in \{x_1, \ldots, x_n, y_1, \ldots, y_{k-1}\}\setminus\{x_l\}$} 
    \nonumber \\
    y_{m+s}  = x_{l_s} \cdot z_w 
    & & \mbox{for $p = 1 \ldots, h$ and $s = 0 \ldots, r_{x_l}'$}
\end{eqnarray}
\begin{eqnarray}
    \lefteqn{y_k - \sum_{s=0}^{r_{x_l}'} v_s^2 \cdot y_{m+s} - 
    \sum_{s=0}^{r_{x_l}'-1} \sum_{t=s+1}^{r_{x_l}'}
    2 \cdot v_s \cdot v_t \cdot y_{m+r_{x_l}'+\eta} = 0} \\ 
    & & \mbox{if $z_v = z_w = x_l$, $\eta = \sum_{\sigma=0}^s \sum_{\theta=\sigma+1}^t 
    1$} \nonumber
\end{eqnarray}
\begin{eqnarray}
    y_{m+s} = x_{l_s}^2 & & \mbox{for $s=0, \ldots, r_{x_l}'$}\\
    y_{m+r_{x_l}'+\eta} = x_{l_s} \cdot x_{l_{t}} 
    & & \mbox{where  $\eta = \sum_{\sigma=0}^s \sum_{\theta=\sigma+1}^t 1$}
\end{eqnarray}
\begin{eqnarray}
    x_i \in D(x_i) \subset \mathbb{Z} & & \mbox{for $i = 1, \ldots, l-1, l+1, \ldots n$} 
    \\
    y_k \in D(y_k) \subset \mathbb{Z} & & \mbox{for $k = 1, \ldots, m + \ldots$} \\
    x_{l_s} \in \{0,1\} & & \mbox{for $s=0,\ldots, r_{x_l}'$}
\end{eqnarray}
The set of substitutions~$S$ is extended accordingly:
\begin{eqnarray}
    S & := & S \cup \{x_l := \sum_{s=0}^{r_{x_l}'} v_s \cdot x_{l_s} \} \enspace.
\end{eqnarray}

The direct substitution of any~$y_l$ with $l \in \{1, \ldots, m\}$ can be avoided if the
variables $z_v, z_w \in \{x_1, \ldots,x_n\}$ with $y_1 = z_v \cdot z_w$ are substituted; 
see Section~\ref{sec:canonical} for details.

There are $r_{x_l}'$ new binary variables introduced and either zero or $r_{x_l}'$ or at 
most $r_{x_l}'+r_{x_l}'(r_{x_l}'-1)/2$ new integer variables. It is recommended to 
substitute integer variables with rather large domains, because the number of introduced 
binary variables will then be less than the number of variables using one hot encoding.

Also, this encoding requires only a finite number of transformation
steps until all variables are binary.

\subsection{Transforming Binary Products in QUBO Form}\label{sec:binprod2qubo}

Let two binary variables $x$ and $y$ with domains $D_x = \{0,1\}$ and $D_y = \{0,1\}$ be 
given. The \emph{binary product constraint} $z = x \cdot y$ is logically equivalent to 
\begin{eqnarray}
    z & \leftrightarrow & x \land y\label{eqn:binproduct}
\end{eqnarray}
if we consider binary variables as Boolean variables. Both can be further represented
by the Rosenberg quadratization \emph{penalty 
term}\footnote{\href{https://docs.dwavesys.com/docs/latest/handbook\_reformulating.html\#cb-techs-reduction-sub-bool}{https://docs.dwavesys.com/docs/latest/handbook\_reformulating.html\#cb-techs-reduction-sub-bool}} 
$x \cdot y - 2 \cdot (x + y )\cdot z +3\cdot z$, i.e. $z = x \cdot y$ is satisfied
if and only if
\begin{eqnarray}
    x \cdot y -2 \cdot x \cdot z - 2\cdot y \cdot z + 3 \cdot z & = &  0
\end{eqnarray}
holds. Further, it always holds that
\begin{eqnarray}
    x \cdot y -2 \cdot x \cdot z - 2\cdot y \cdot z + 3 \cdot z & \ge &  0 \enspace.
\end{eqnarray}
Justification:
We consider two cases: a) if either $x$ or $y$ is zero the term~$x \cdot y -2 \cdot x 
\cdot z - 2\cdot y \cdot z + 3 \cdot z$ reduces to $3 \cdot z$ which is zero if $z$ is
zero and greater than zero (namely 3) if $z = 1$. b) when $x = y = 1$ the term reduces 
to $1 - z$ which becomes zero if $z=1$ and one if $z=0$. Thus the product 
$x \cdot y = z$ is satisfied if and only if the QUBO sub-problem
\begin{eqnarray}
    q_{\fs{prod}}(x,y,z) 
    & \equiv & x \cdot y -2 \cdot x \cdot z - 2\cdot y \cdot z + 3 \cdot z 
    \label{expr:QUBOproduct}
\end{eqnarray}
is minimised. 

\subsection{Transforming Binary QIP(FD) into QUBO Problems}\label{sec:toQUBO}

We assume that there is a binary QIP(FD) defined by the Equations~(\ref{eqn:orgObj}) -- 
(\ref{eqn:orgLast}) where $p=0$ holds, i.e. without Inequalities~(\ref{eqn:orgIneqs})
be given. We show how to transform this QIP(FD) into a QUBO problem.

Let a binary QIP(FD) defined by the Equations~(\ref{eqn:orgObj}) -- (\ref{eqn:orgLast}) 
where $p=0$ holds, i.e. without Inequalities~(\ref{eqn:orgIneqs}) be given.
Then the resulting QUBO is 
\begin{eqnarray}
    \lefteqn{\min \Bigg(\sum_{i=1}^n g_i \cdot x_i + \sum_{j=1}^q  C_j \cdot 
    \Big(\sum_{i=1}^n d_{j,i} \cdot x_i + \sum_{k=1}^m e_{j,k} \cdot y_k + f_j\Big)^2}
    \label{eqn:squaring}\\
    & & \qquad +~\sum_{k=1}^m 
    C_k \cdot \big(z_v \cdot z_w - 2\cdot z_v \cdot y_k - 2\cdot z_w \cdot y_k + 3 \cdot 
    y_k\big) \Bigg) \enspace. \\[-2ex]
    & & \qquad\qquad\qquad
    \mbox{where $z_v, z_w \in \{x_1, \ldots, x_n, y_1, \ldots, y_{k-1}\}$} 
    \nonumber  
\end{eqnarray}
Please note that squaring of the sums in Term~(\ref{eqn:squaring}) is not necessary 
if these terms cannot become less than zero. This can be decided by the use of 
the bounds of the variables domains. By the way:  If the lower bound of one of
these terms is greater than zero then the according constraint cannot be satisfied,
i.e. the whole problem is inconsistent.

Finally the resulting QUBO can be solved by the use of a Simulated Annealer, a 
Quantum-inspired Annealer or a Quantum Annealer. Then, the values of the `original'
integer variables and thus the result of the `original' QIP(FD) can be computed
by evaluating the equations in~$S$.

\section{Conclusion and Future Work}

We presented all transformation steps which are necessary to transform any
QIP(FD) into an equivalent QUBO which can be solved on the Quantum Annealer. Furthermore,
we maintain the substitutions which are necessary the compute the solutions of the
transformed QIP(FD) from the solution of the QUBO. Based on these results
we implemented converters which translates {\tt FlatZinc} into QIP(FD) programs.
These QIP(FD) programs are represented by data structures of the selected programming
language. In detail we implemented a workflow performing of the following tasks:
\begin{itemize}
    \item introduce \emph{slack variables} to remove the linear inequalities 
        (cf.~Section~\ref{sec:slack}),
    \item perform \emph{bounds consistency} (cf.~\cite{schulteWhenBoundsDomain2001} 
        to reduce the domains of the variables,
    \item make the domains of the variables canonical (cf.~Section~\ref{sec:canonical})
    \item encode the integer variables by binary variables 
        (cf.~Sections~\ref{sec:oneHotEncoding} and~\ref{sec:binEncoding})
    \item transfer the binary QIP(FD) into a QUBO (cf.~Sections~\ref{sec:factors},
        \ref{sec:binprod2qubo} and~\ref{sec:toQUBO}) which will be the input of
        Quantum Annealer.
\end{itemize} 
Our future work will focus on testing and improving this MiniZinc-to-QUBO 
converter on sample MiniZinc programs like the SEND-MORE-MONEY problem or on 
job-shop-scheduling problems.

\bibliographystyle{plain}
\bibliography{references}
\end{document}